\documentclass{aa}
\usepackage{graphicx}
\usepackage{natbib}
\usepackage{txfonts}
\usepackage{color}

\def\ros{{\it ROSAT}}
\def\chandra{{\it Chandra}}

\def\xmm{{\it XMM-Newton}}
\def\n253{NGC~253}
\def\m82{M82}
\def\deml71{DEM~L~71}
\def\la{\mathrel{\hbox{\rlap{\hbox{\lower4pt\hbox{$\sim$}}}\hbox{$<$}}}}
\def\ga{\mathrel{\hbox{\rlap{\hbox{\lower4pt\hbox{$\sim$}}}\hbox{$>$}}}}

\def\deg{\hbox{$^\circ$}}

\def\cm-2{cm$^{-2}$}

\newcommand{\nh}{\hbox{$N_{\rm H}$}}

\newcommand{\ct}{ct~s$^{-1}$}

\newcommand{\ie}{i.e.\ }

\begin{document}
   \title{High-resolution X-ray spectroscopy and imaging of the nuclear outflow of the starburst galaxy \n253\thanks{Based on observations obtained with XMM-Newton, an ESA science mission with instruments and contributions directly funded by ESA Member States and NASA}}

   \author{M.~Bauer\inst{1}, W.~Pietsch\inst{1},  G.~Trinchieri\inst{2}, D.~Breitschwerdt\inst{3}, M.~Ehle\inst{4} \and  A.~Read\inst{5} 
          }
   \authorrunning{M. Bauer et al.}

   \offprints{M. Bauer, \email{mbauer@mpe.mpg.de}}

   \institute{Max-Planck-Institut f\"ur extraterrestrische Physik, Giessenbachstra\ss e, 85741 Garching, Germany
   \and
   INAF Osservatorio Astronomico di Brera, via Brera 28, 20121 Milano, Italy
   \and
   Institut f\"ur Astronomie der Universit\"at Wien, T\"urkenschanzstr. 17, A-1180 Wien, Austria
   \and
   European Space Astronomy Centre, ESA, P.O. Box 50727, 28080 Madrid, Spain
   \and
   Department of Physics \& Astronomy, University of Leicester, Leicester LE1 7RH, UK
    }

   \date{Received 4 September 2006; accepted 18 January 2007.}


  \abstract
   {}
   {Using \xmm\ data, we have aimed to study the nuclear outflow of the nearby starburst galaxy \n253\ in X-rays with respect to its morphology and to spectral variations along the outflow.
   }
   {We analysed \xmm\ RGS spectra, RGS brightness profiles in cross-dispersion direction, narrow band RGS and EPIC images and EPIC~PN brightness profiles of the nuclear region and of the outflow of \n253.
   }
   {We detect a diversity of emission lines along the outflow of \n253.
   This includes the He-like ions of Si, Mg, Ne and O and their corresponding ions in the next higher ionisation state.
   Additionally transitions from \ion{Fe}{XVII} and \ion{Fe}{XVIII} are prominent.
   The derived temperatures from line ratios along the outflow range from $0.21\pm0.01$ to $0.79\pm0.06$~keV and the ratio of \ion{Fe}{XVII} lines indicates a predominantly collisionally ionised plasma.
   Additionally we see indications of a recombining or underionized plasma in the \ion{Fe}{XVII} line ratio.
   Derived electron densities are $0.106\pm 0.018$~cm$^{-3}$ for the nuclear region and $0.025\pm 0.003$~cm$^{-3}$ for the outflow region closest to the centre.
   The RGS image in the \ion{O}{VIII} line energy clearly shows the morphology of an outflow extending out to $\sim$750~pc along the south-east minor axis, while the north-west part of the outflow is not seen in \ion{O}{VIII} due to the heavy absorption by the galactic disc.
   This is the first time that the hot wind fluid has been detected directly.
   The limb brightening seen in \chandra\ and \xmm\ EPIC observations is only seen in the energy range containing the \ion{Fe}{XVII} lines (550--750~eV).
   In all other energy ranges between 400 and 2000~eV no clear evidence of limb brightening could be detected.
   }
   {}

   \keywords{X-rays: galaxies -- Galaxies: individual: \n253 -- Galaxies: spiral -- Galaxies: starburst -- ISM: jets and outflows}

   \maketitle


\section{Introduction}
Starburst galaxies are known to show very complex emission in X-rays.
This emission originates on the one hand from sources that appear to be point-like sources, like X-ray binaries, supernovae and supernova remnants.
On the other hand emission comes from the diffuse hot component of the interstellar medium like diffuse emission in the disc and gaseous outflows driven out of the disc by massive stellar winds and core collapse supernovae, also called superwinds.
The latter phenomenon can be quite spectacular in \m82\ \citep[e.g.][]{SR2003} and \n253\ \citep[e.g.][]{SH2000}, for example, where these superwinds emerge from a starburst nucleus.
In \n253, \cite{FT1984} first detected the south-east part of this outflow in X-rays with {\it Einstein} and called it the "minor-axis component".
With \ros, \cite{PV2000} also detected the part of the outflow pointing into the opposite direction.
However the spatial resolution of \ros\ was not yet good enough to learn more about the morphology of this outflow.
Later on, observations with \xmm\ \citep{PR2001} and especially with \chandra\ \citep{SH2000} showed that the outflow can be explained with a limb brightened hollow cone structure.
Temperatures of the best fit thin thermal plasma models are in the range 0.15-0.94~keV from \xmm\ EPIC and 0.46-0.66~keV from \chandra.
\cite{SH2000} concluded that the detected emission originates from the shocked region at the border of the outflow where the wind collides with interstellar medium.
The wind itself though was thought to be too hot and too thin to be detected directly. 
This picture, however, disregards the possibility that the wind may be mass-loaded, 
entraining ambient interstellar medium (ISM) as well as infalling material. If turbulent 
mixing proceeds on a time scale which is larger than the flow time within a 
given region, such as the base of the outflow studied here, we expect some 
clumpiness in the outflow, imprinted on an overall \textit{less dense} wind. 
As we will show later, this is confirmed by our analysis, which shows that e.g. \ion{O}{VIII} is \textit{not} 
limb brightened.

High resolution spectra of \n253\ and \m82, taken with the \xmm\ Reflection Grating Spectrometer (RGS) were first published by \cite{PR2001} and \cite{RS2002}, respectively.
Both spectra show the Ly$_{\alpha}$ emission lines from Si, Mg, Ne, O, N and also their helium-like charge states.
Also both galaxies show emission lines from \ion{Fe}{XVII} and \ion{Fe}{XVIII}, and \m82\ additionally shows lines from Fe~XX, Fe~XXIII and Fe~XXIV.
In \m82\ the line ratios for neon, iron and oxygen are quite different compared to \n253.
In general the \m82\ spectrum appears to be hotter with temperatures in the range of $\sim$0.3--1.5~keV, with its continuum more confined to higher energies.
Its X-ray flux, as well as its X-ray luminosity, in the RGS energy band (0.35--2.5~keV) is higher than that of \n253.

However these spectra only give a combined spectrum of the nuclear source and the outflow.
In this paper we present an analysis where we decompose the total spectrum of \n253\ into regions containing the nucleus and different parts of the outflow, while maintaining the high spectroscopic resolution.


\section{Observations and data reduction}

The nuclear region of \n253 was observed with \xmm\ \citep{JL2001} during two orbits in June 2000 and June 2003 using all of the European Photon Imaging Camera (EPIC) instruments \citep{SB2001,TA2001} and the two co-aligned RGS spectrometers, RGS1 and RGS2 \citep{HB2001}, for a total of about 190~ks.
The details of these observations are shown in Table~\ref{tab:Exposures}.
An additional archival observation (Obs. id. 0110900101) could not be used for this analysis, since the pointing of this observation was in the north-west halo of \n253 with the result that the outflow was not in the field of view (FOV) of the RGS.

\begin{figure}
   \centering
   \includegraphics[width=\columnwidth]{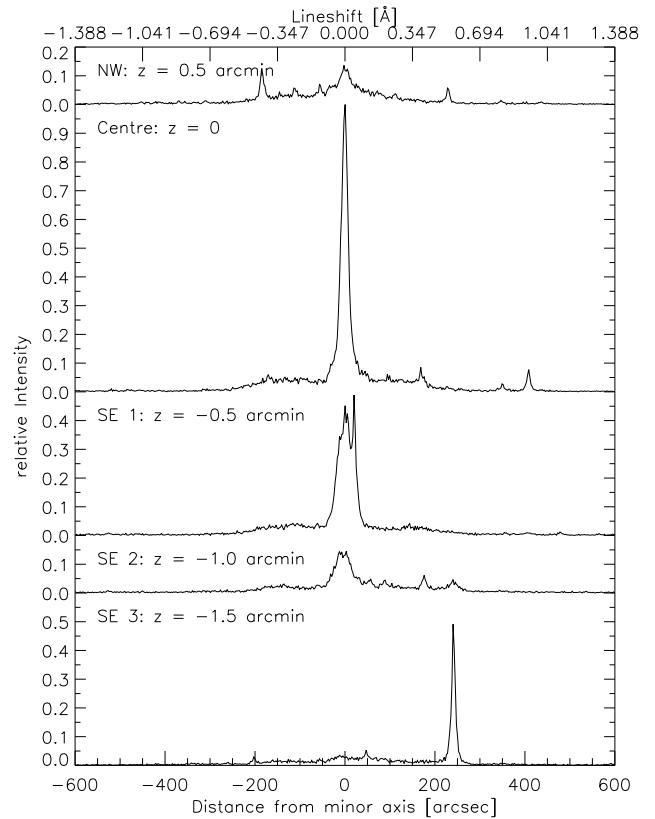}
   \caption{EPIC~PN brightness profiles along the RGS dispersion direction of each extraction region. 
   The strong peak at 0\arcsec\ is caused by the outflow emission. 
   In the Centre region, shown in Fig.~\ref{fig:RGSRegions}, this is superimposed by the nuclear source of the galaxy. 
   Point sources in the extraction regions are seen as sharp spikes in the profile.
   A positive distance points parallel to the major axis towards the south-west.
   The distance from the galactic major axis is given by the value $z$ in the captions of the individual extraction regions.}
   \label{fig:BrightnessProfile}
\end{figure}

Before we start to describe the analysis procedures, we want to place emphasis on why it is possible at all to perform the following spectroscopic analysis.
First, the nuclear outflow of \n253\ is an extended X-ray object, which can be spatially resolved by \xmm\ EPIC and RGS, embedded in an even larger region of X-ray emission from point-like sources and diffuse emission in the disc and halo of the galaxy.
This means the RGS data for the central regions are in principle affected by the contamination from the surrounding
emission, but, as shown in Fig.~\ref{fig:BrightnessProfile}, both the nucleus and the outflow in \n253\ are significantly brighter and well above the galaxy emission, so we can expect only a minor contamination.
Moreover as we show later, we can identify and "remove" effects due to the disc emission.

Second, since the RGS is a slitless spectrometer, the spectra of all sources in the field of view are superimposed on each other on the detector.
A spatial displacement of a source along the dispersion direction corresponds to a wavelength shift in the spectrum  of $2.31\times 10^{-3}$~\AA~arcsec$^{-1}$ with respect to a not-displaced source.
Since the outflow has an extent of up to 1.4\arcmin\ in the dispersion direction, the spectral resolution is limited to $\sim$0.19~\AA\ at 15~\AA.
This is still considerably better than the energy resolution from CCD detectors.

\begin{figure*}
   \centering
   \includegraphics[width=\textwidth]{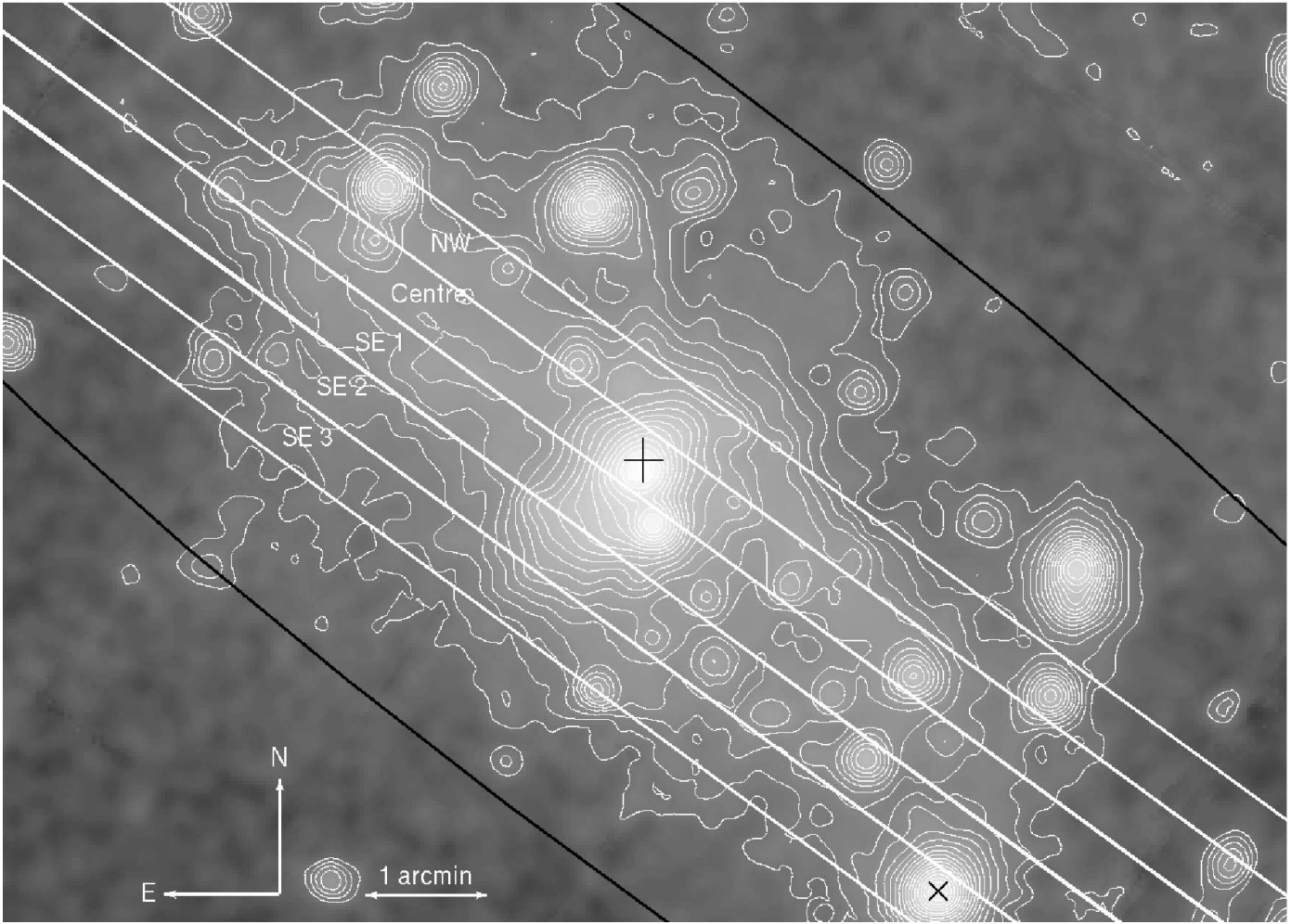}
   \caption{Logarithmically scaled EPIC~PN+MOS image of the central region of \n253 in the energy band $0.5-2.0$~keV.
   The centre of the galaxy is marked with the black cross.
   The D25 ellipse is overlaid in black, brightness contours and the extraction regions for the RGS spectra are overlaid in white.
   The black X at the bottom of the image marks the source X21.
   }
   \label{fig:RGSRegions}
\end{figure*}

Due to the superposition of all sources, other bright point sources in the FOV could contaminate the spectrum of the outflow.
However the effective area decreases significantly for off-axis sources and even a contribution from the brightest off-axis source at the bottom of Fig.~\ref{fig:RGSRegions} \citep[X21 from ][]{PV2000,TS2005} can be neglected.
The source at the south-west edge of the outflow in region SE~1 (cf.\ Fig.~\ref{fig:BrightnessProfile} and \ref{fig:RGSRegions}) however does affect the outflow spectrum.
The spectrum of this source does not show line features \citep[X33 in][]{PR2001}, so its contribution to the RGS spectrum from this region is an increased continuum flux.
This does not affect our conclusions.

\begin{table*}
\caption{\xmm\ \n253\ observation log. Observation identification, observing date, pointings and orientation of the satellite, total exposure time (T$_{\rm{exp}}$) and exposure time after screening for high background (T$_{\rm{exp,\ clean}}$) are given.}
\label{tab:Exposures}
\begin{center}
\begin{tabular}{ccccccccccc}
\hline\noalign{\smallskip}
\hline\noalign{\smallskip}
\multicolumn{1}{c}{Nr.} & \multicolumn{1}{c}{Obs. id.} &\multicolumn{1}{c}{Obs. dates} & \multicolumn{2}{c}{Pointing direction} & \multicolumn{1}{c}{P. A.}& \multicolumn{1}{c}{T$_{\rm{exp}}$}& \multicolumn{1}{c}{T$_{\rm{exp,\ clean}}$}\\
\noalign{\smallskip}
 & & & \multicolumn{2}{c}{RA/DEC (J2000)} & \multicolumn{1}{c}{(deg)}& \multicolumn{1}{c}{(ks)}& \multicolumn{1}{c}{(ks)}\\
\noalign{\smallskip}
\multicolumn{1}{c}{(1)} & \multicolumn{1}{c}{(2)} & \multicolumn{1}{c}{(3)} & \multicolumn{1}{c}{(4)} & \multicolumn{1}{c}{(5)} & \multicolumn{1}{c}{(6)} & \multicolumn{1}{c}{(7)} & \multicolumn{1}{c}{(8)} \\
\noalign{\smallskip}
\hline
\noalign{\smallskip}
1 & 0125960101 & 2000-06-03 & 00:47:36.74 & -25:17:49.2 & 56.9 & 60.8  & 45.1 \\
2 & 0125960201 & 2000-06-04 & 00:47:36.57 & -25:17:48.7 & 57.0 & 17.5  & 7.0  \\
3 & 0152020101 & 2003-06-19 & 00:47:36.89 & -25:17:57.3 & 53.8 & 113.0 & 75.9 \\
\hline
\end{tabular}
\end{center}
\end{table*}

We analysed the data using the Science Analysis System ({\tt SAS}), version 6.5.0, together with the most recent calibration files available at the time of the analysis. The metatask {\tt rgsproc 1.19.6} was used to process the RGS data.
We first filtered the data sets for time periods with low contamination by low energy protons. 
Then we extracted a light curve from the background region on CCD~9, the chip closest to the optical axis of the telescope, therefore the most affected by background flares, to determine the threshold count rate, which we then used to filter the eventfiles. 
The thresholds were 0.25~\ct\ for observations 0125960101 and 0125960201, and 0.20~\ct\ for observation 0152020101, where the count rate was more stable during the non-flaring time intervals.

\subsection{RGS spectra}
Since the dispersion direction of the RGS was aligned approximately with the major axis of the galaxy we were able to extract spectra for different adjacent regions along the minor axis, \ie the cross-dispersion direction, of the galaxy (see Fig.~\ref{fig:RGSRegions}).
The extent of the extraction regions are $30\arcsec$ in cross-dispersion direction.
Assuming a distance to \n253 of 2.58~Mpc \citep{PC1991} this corresponds to a width of the extraction regions of 375~pc.
The events in these regions were additionally filtered with a CCD pulseheight filter to select only the $m=-1$ spectral order.
The reference points for the origins of the energy scales of the spectra were set on the minor axis of the galaxy, where the outflow has its peak emission (see Fig.~\ref{fig:BrightnessProfile}).

\n253\ is an extended source and covers most of the area of the RGS detectors.
To prevent contamination in the background spectra we used the task {\tt rgsbkgmodel 1.1.5}, which computes background spectra from RGS background templates.

To increase statistics we combined the spectra of the two RGS detectors and added up the spectra from the three observations.
As the position angles of \xmm\ did not differ very much in the three observations, the regions from which the spectra were extracted are only slightly tilted relative to each other.
A difference in position angle between observations has the effect of degrading the spatial coincidence in the extraction regions.
However in our case position angle differences are small enough that we can neglect this error.
A correction of the spectra for effective area and the combination of the spectra from different observations and instruments was done with the task {\tt rgsfluxer}. 
The task's description states that the fluxed spectrum produced by {\tt rgsfluxer} should not be used for any serious analysis of the data.
If we use it in spite of this warning, we have to consider the following effect that will add to uncertainties:
The task {\tt rgsfluxer} neglects the redistribution of monochromatic response into the dispersion channels, so the intrinsic line broadening of the detector is not removed from the spectrum.
As we do not determine line positions or line widths and since we integrate over the whole line including its wings to derive fluxes for individual lines, this effect does not restrict our analysis.
Additionally we are only interested in relative line fluxes and not in absolute values. 

To obtain acceptable statistics ($>3\sigma$) for most of the emission lines while maintaining a high spectral resolution of $\sim$0.39~\AA\ in the spectra we combined six channels at a time.

For the observations the dispersion direction of the spectrometers was approximately aligned along the major axis of the galaxy.
Therefore the spectra for different cross-dispersion areas correspond to regions with different distance $z$ from the galactic plane (cf. Fig.~\ref{fig:RGSRegions}).
The spectra obtained from these regions are shown in Fig.~\ref{fig:spec}.

\subsection{RGS images}
The RGS is a slitless spectrometer.
So the whole observed target within the FOV is imaged on the detector plane, with an offset in dispersion direction for every wavelength it emits in.
This aspect can be used to extract narrow band images for various emission lines.
In cross-dispersion direction the image in a selected emission line is directly mapped onto the detector CCD, while in dispersion direction the observed object is compressed into a narrow region.
The technique was first applied to RGS data of the supernova remnant \deml71 by \cite{HB2003}.
With the help of K.~J. van der Heyden (private communications) we developed our own code to produce these narrow band images.
The procedure is as follows:
The low background eventlists are filtered for the wavelength range of the desired line and the "banana region" in wavelength-energy-space to exclude second order spectra and noise.
By setting the wavelength range narrow enough we made sure that no neighbouring lines would be in the extraction region.
The dataset thus obtained has to be converted into spatial coordinates and it has to be uncompressed along the dispersion axis using the following equation as described in the SAS task {\tt rgsangles}:
\begin{equation}
\Delta \beta=\frac{\sin\alpha}{\sin \beta}\Delta\phi\frac{F}{L}
\end{equation}
with the change in the grating exit angle $\Delta\beta$ due to the offset in the angular component $\Delta\phi$ of an off-axis source parallel to the dispersion direction, the angle of incidence $\alpha$, the grating exit angle $\beta$, the focal length $F$, and the distance between the Reflection Grating Assembly and the prime focus $L$.

The images were corrected for exposure and binned to a pixel size of 0.4\arcsec.
In a next step we included a RA-DEC coordinate system.
As reference coordinate, we chose the coordinate of the centre of \n253\ \citep[the position of the brightest IR source in the galaxy, $\alpha_{2000}=0^{\rm h} 47^{\rm m}$ 33\fs 3, $\delta_{2000}=-25$\deg 17\arcmin 18\arcsec,][]{FP2000}.
In cross-dispersion direction the position of our reference coordinate on the CCD could be taken directly from the source list file which was produced in the processing of the RGS data.
In dispersion direction the position is given in the above procedure by the Doppler shift corrected line centre position.
The velocity that had to be accounted for is the galaxy's systemic velocity of 243~km~s$^{-1}$ \citep{KS2004}.
This shifts the reference coordinate along the dispersion direction by $\sim$0.35\arcsec~\AA$^{-1}$ times the centre wavelength in which the image is calculated, e.g.\ 6.6\arcsec\ for the \ion{O}{VIII} image.
The effects of the velocities of the earth with respect to the sun, and \xmm's orbital velocity can be neglected since they are only of the order of 0.7\arcsec\ and 4\arcsec$\times 10^{-4}$, respectively in the \ion{O}{VIII} image where the effect would be largest, and therefore much smaller than the width of the point spread function.
The images for the lines were created separately for each of the three observations and were then combined into one image.
In a final step the images were smoothed with a gaussian filter.

In general the method is affected by two different effects:
(i) A Doppler shift due to the radial velocity component of an emitting source changes its position in the image along the dispersion direction axis. 
A radial velocity of 1000~km~s$^{-1}$ would correspond to 21.6\arcsec at a wavelength of 15.0~\AA. 
(ii) Assuming there is not only one but two lines in the wavelength extraction interval, we would have two images of the object in the resulting image, superimposed with an offset along the dispersion direction of 7.2\arcmin~\AA$^{-1}$. For example the emission in the two \ion{Fe}{XVII} lines at 16.780~\AA\ and 17.055~\AA\ would be superimposed with an offset of $\sim$2.0\arcmin.
The \ion{Fe}{XVII} at 17~\AA\ image is the only case where we actually have to consider that we have created an image using two lines, \ie the \ion{Fe}{XVII} lines at 17.055~\AA\ and at 17.100~\AA.
In the spectra the lines are not separated and they appear about equally strong, which is most likely an effect due to the low statistics.
According to theory \citep{MG1985} the line strength of the latter line should be $\sim$52--85~\% of the former one, depending on the temperature of the plasma (1.08~keV and 0.11~keV, respectively).
The separation of the lines is 0.046~\AA\ which corresponds to a shift in dispersion direction of 19.9\arcsec.
By smoothing the image with a larger gaussian with a FWHM of 20\arcsec\ we can account for the error we make by using both lines.

\begin{figure*}
   \centering
   \includegraphics[width=14cm]{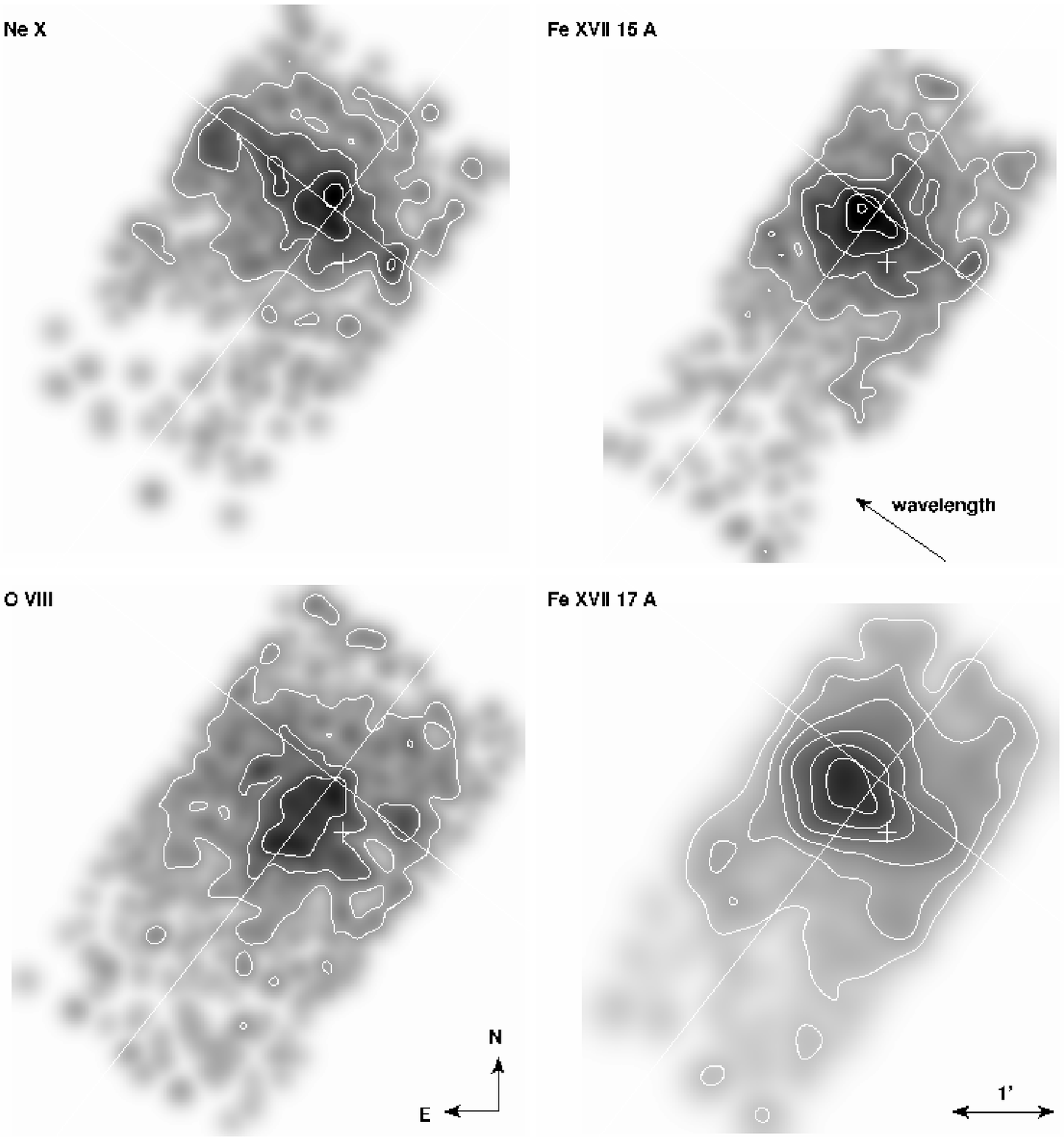}
   \caption{RGS images of \n253\ in the \ion{Ne}{X}, \ion{Fe}{XVII} and \ion{O}{VIII} lines.
   The white lines mark the major (north-east to south-west) and minor axis (north-west to south-east) of \n253.
   The image in the \ion{Fe}{XVII} line at 17~\AA\ was smoothed with FWHM of 20\arcsec, all others with 12\arcsec.
   White contours indicate 2$\sigma$, 3$\sigma$, 4$\sigma$, \dots above the background.
   The white cross south-west of the nucleus marks the bright source X33 from \cite{PV2000}.
   The RGS dispersion direction is such that wavelength increases from south-west to north-east as indicated by the arrow.
   }
   \label{fig:RGSNeFeO}
\end{figure*}

Images in the \ion{Ne}{X}, \ion{Fe}{XVII} and \ion{O}{VIII} lines are shown in Fig.~\ref{fig:RGSNeFeO}.
The number of photons that were extracted from all three observations are 806, 1077, 816 and 1231 for the images in the \ion{Ne}{X} (11.98--12.35~\AA), \ion{Fe}{XVII} at 15~\AA\ (14.86--15.13~\AA), \ion{Fe}{XVII} at 17~\AA\ (16.90--17.21~\AA) and \ion{O}{VIII} (18.80--19.17~\AA) lines respectively.
The FWHM of the gaussian filter was 12\arcsec\ for \ion{Ne}{X}, \ion{Fe}{XVII} at 15~\AA, \ion{O}{VIII}, and 20\arcsec\ for \ion{Fe}{XVII} at 17~\AA.

\subsection{RGS cross-dispersion profiles}
To get additional spatial information of the line distribution, we produced emission line profiles in the cross-dispersion direction.
Therefore we extracted events from the RGS eventfiles by applying the same filters in wavelength and wavelength-energy-space as for generating the RGS images, but then we binned the counts into 30\arcsec\ bins to match the extraction regions that were applied for the spectra. 
Background counts in the respective wavelength ranges were taken from the spectra that were obtained with the RGS background model task and subtracted from the emission line profiles.
Four of these profiles are shown in Fig.~\ref{fig:RGSprofilesAll}.

\subsection{EPIC-PN images}
To verify the results from the RGS images we also extracted EPIC~PN narrow band images in approximately the same energy ranges.
Therefore we filtered the PN eventfiles in the energy bands around \ion{Ne}{X} (992--1052~eV), \ion{Fe}{XVII} at 15~\AA\ (795--844~eV), \ion{O}{VIII} (625--690~eV) and \ion{Fe}{XVII} at 17~\AA\ (694--734~eV).
The spectral resolution of the EPIC~PN detector is $\sim$70~eV, so it is possible that photons with higher or lower energies contribute to the energy band of interest.
Additionally there is contamination from higher energies due to the redistribution in the detector.
Photons can lose up to 60\% of their energy in the CCD before they are detected.
This means that bright features in some energy range can show up to some degree in lower energy bands.
The filtered eventfiles of the different observations and instruments were merged using the {\tt SAS} task {\tt merge}.
We created images of the eventfiles and smoothed them with a gaussian of 6\arcsec.
The resulting images are shown in Fig.~\ref{fig:EPICNeFeO}.

\subsection{EPIC-PN brightness profiles}
To detect the limb brightening of the outflow, as found by \cite{SH2000} and \cite{PR2001}, we also extracted brightness profiles perpendicular to the outflowing direction from the merged PN eventfiles of all observations.
Furthermore to check for an energy dependence in the limb brightening we subdivided them into energy bins with a width of $\sim$150~eV starting from 400~eV up to 2000~eV.
The emission lines of \ion{O}{VIII} and the \ion{Fe}{XVII} lines are included in the energy ranges 550--700~eV and 700--850~eV, respectively.
\ion{Ne}{X} is mostly in the 1000-1150~eV bin.
The extraction regions match the regions we used for the RGS spectra.
Furthermore we split the region SE~1 into the two regions 'SE~1 (1)' and 'SE~1 (2)' with a width of 15\arcsec\ in cross-dispersion direction of the RGS.
The brightness profiles, sorted for energy band, are shown in Fig.~\ref{fig:PN_profiles}.
Since we are not interested in the emission from the nucleus nor from the bright source X33 south-west of the nucleus, the profiles are limited to a maximum of 250 counts.
This still shows the main features in the Centre and SE~1 (1) region, while cutting off the peaks in some cases.


\section{Results}
\subsection{RGS spectra}
\begin{figure*}
\centering
\includegraphics[width=18cm]{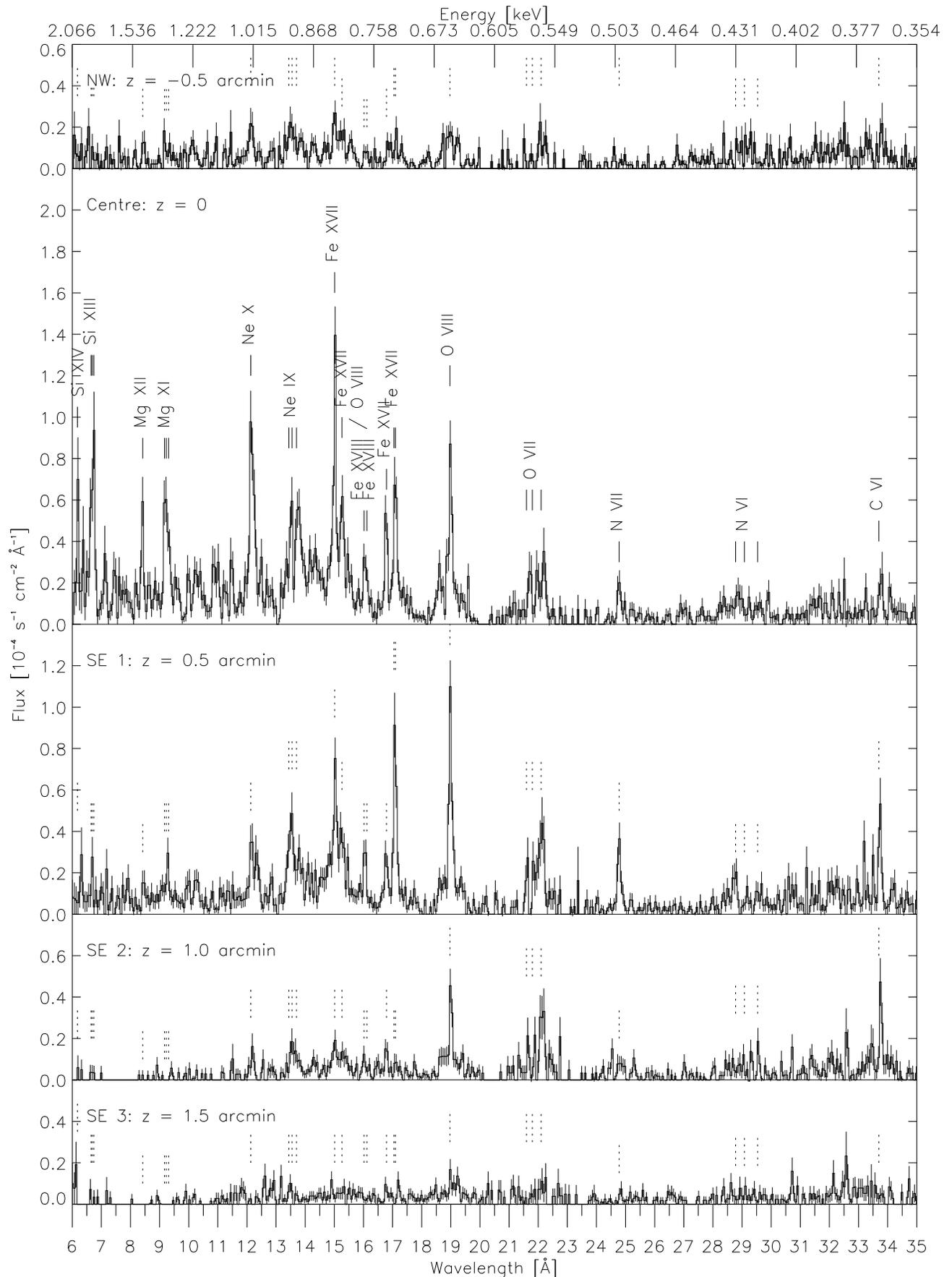}
\caption{Combined RGS spectra of \n253\ extracted from different regions along the outflow.
The label gives the south-east position z of the extraction region along the minor axis relative to the centre of the galaxy in arcmin.
\label{fig:spec}}
\end{figure*}

\begin{table*}
  \begin{center}
    \caption{Flux values for different emission lines in the extraction regions. The references for the expected wavelengths ($\lambda_{\mathrm{expected}}$) are \cite{MG1985} and \cite{PM1999}.}
    \vspace{1em}
    \begin{footnotesize}
    \renewcommand{\arraystretch}{1.2}
    \begin{tabular}[h]{lccccc}
\hline\noalign{\smallskip}
\hline\noalign{\smallskip}
Transition  & $\lambda_{\mathrm{expected}}$ & \multicolumn{4}{c}{Flux}               \\
&(\AA)&\multicolumn{4}{c}{(10$^{-6}$ s$^{-1}$ cm$^{-2}$)}\\
\hline                    
  &  & NW   & Centre   & SE 1   & SE 2   \\
\hline                    
\ion{Si}{XIV}  			& 6.18 &    								& 5.6 $\pm$ 4.3 &    							&    								\\
\ion{Si}{XIII} w 		& 6.65 &   									& 5.8 $\pm$ 3.4 &   					 		&    								\\
\ion{Si}{XIII} x+y 	& 6.69 &    								& 4.0 $\pm$ 2.8 &    							&    								\\
\ion{Si}{XIII} z 		& 6.74 &    								& 9.2 $\pm$ 4.3 &    							&    								\\
\ion{Mg}{XII}  			& 7.11 & 								 		& 6.5 $\pm$ 3.0 & 1.9 $\pm$ 1.6 & 								  \\
\ion{Mg}{XI} w 			& 7.76 & 1.4 $\pm$ 1.3   		& 6.5 $\pm$ 2.7 & 1.3 $\pm$ 1.2 &    								\\
\ion{Mg}{XI} x+y 		& 7.81 &    								& 5.4 $\pm$ 2.4 & 1.7 $\pm$ 1.3 &    								\\
\ion{Mg}{XI} z 			& 7.87 &                	 	& 6.3 $\pm$ 2.6 & 2.5 $\pm$ 1.6 & 								 \\
\ion{Ne}{X}  				& 12.1 & 3.4 $\pm$ 2.2	  	& 16  $\pm$ 4.5 & 5.7 $\pm$ 2.6 & 2.0 $\pm$ 1.6 	\\
\ion{Ne}{IX} w 			& 13.5 & 									 	& 4.3 $\pm$ 2.3 & 5.4 $\pm$ 2.5 & 								 	\\
\ion{Ne}{IX} x+y 		& 13.6 & 3.0 $\pm$ 2.0	  	& 6.2 $\pm$ 2.8 & 6.1 $\pm$ 2.7 & 2.4 $\pm$ 1.8 	\\
\ion{Ne}{IX} z 			& 13.7 & 1.8 $\pm$ 1.6 	 		& 7.6 $\pm$ 2.8 & 4.1 $\pm$ 2.2 & 								 	\\
\ion{Fe}{XVII}15 		& 15.0 & 4.5 $\pm$ 1.9 	 		& 18  $\pm$ 3.8 & 13  $\pm$ 3.1 & 3.3 $\pm$ 1.6 	\\
\ion{Fe}{XVII}17  	& 17.1 & 3.0 $\pm$ 1.7	 		& 13  $\pm$ 3.8 & 13  $\pm$ 3.9 & 								 \\
\ion{O}{VIII}  			& 19.0 & 6.7 $\pm$ 2.5 			& 16  $\pm$ 3.8 & 18  $\pm$ 4.0 & 6.9 $\pm$ 2.5 	\\
\ion{O}{VII} w 			& 21.6 &  									& 2.3 $\pm$ 2.2 & 3.0 $\pm$ 2.4 & 								 	\\
\ion{O}{VII} x+y 		& 21.8 &  									& 3.4 $\pm$ 2.6 & 						  & 								 	\\
\ion{O}{VII} z 			& 22.1 & 5.1 $\pm$ 3.3 			& 9.5 $\pm$ 4.4 & 13  $\pm$ 5.2 & 								 	\\
\ion{N}{VII}  			& 24.8 &  									& 2.2 $\pm$ 1.5 & 6.3 $\pm$ 2.8 & 1.5 $\pm$ 1.3	\\
\ion{N}{VI} w 			& 28.8 & 1.4 $\pm$ 1.3 			&    		  			& 4.2 $\pm$ 2.4 &   								\\
\ion{N}{VI} x+y 		& 29.1 & 								 		&    		 				& 2.4 $\pm$ 1.8 &    								\\
\ion{N}{VI} z 			& 29.5 &    								&    		  			& 3.2 $\pm$ 2.0 & 1.8 $\pm$ 1.5 	\\
\ion{C}{VI}  				& 33.7 & 3.0 $\pm$ 2.5	 		& 4.5 $\pm$ 3.0 & 6.6 $\pm$ 3.3 & 5.7 $\pm$ 3.1 	\\
\noalign{\smallskip}
      \hline \\
      \end{tabular}
      \end{footnotesize}
    \label{tab:lineflux}
  \end{center}
\end{table*}

\begin{figure*}
  \hfill
  \begin{minipage}{\textwidth}
   \begin{center}
   \includegraphics[width=5cm]{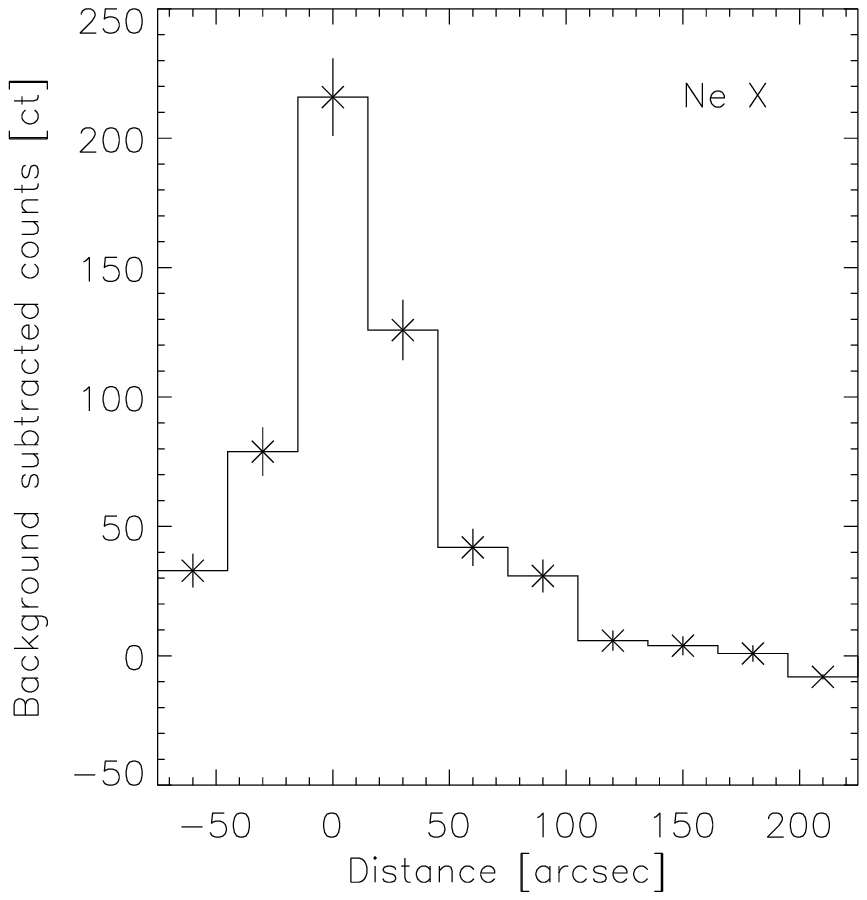}
   \includegraphics[width=5cm]{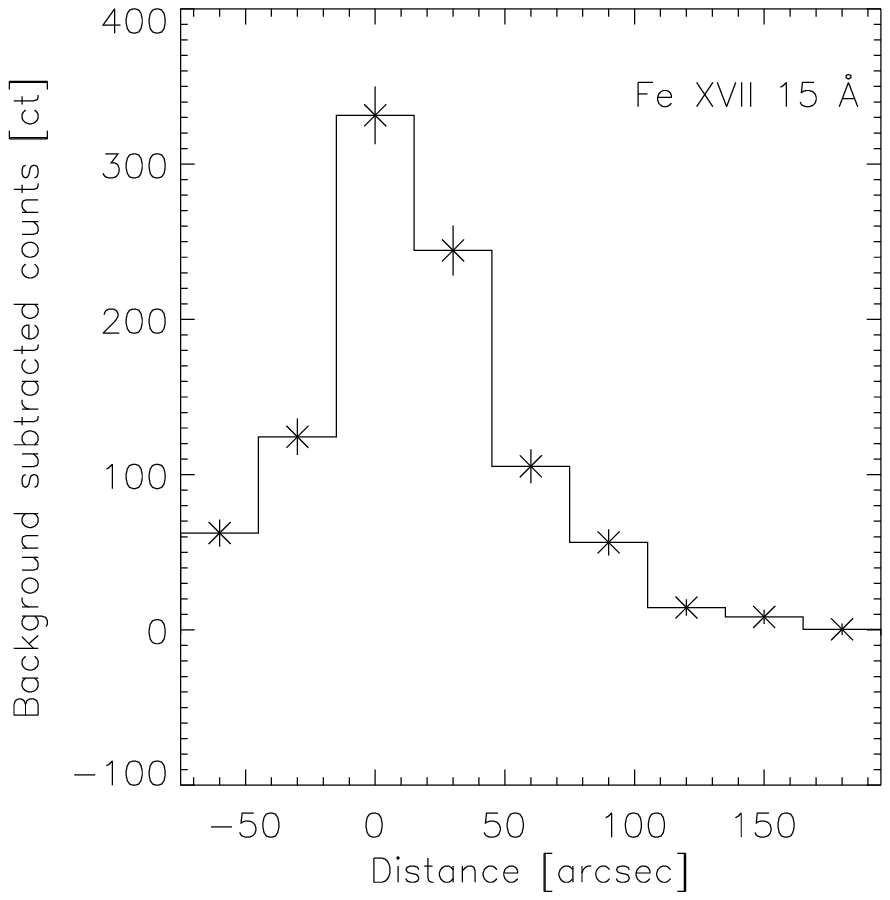}
  \end{center}
  \end{minipage}
  \begin{minipage}{\textwidth}
   \begin{center}
   \includegraphics[width=5cm]{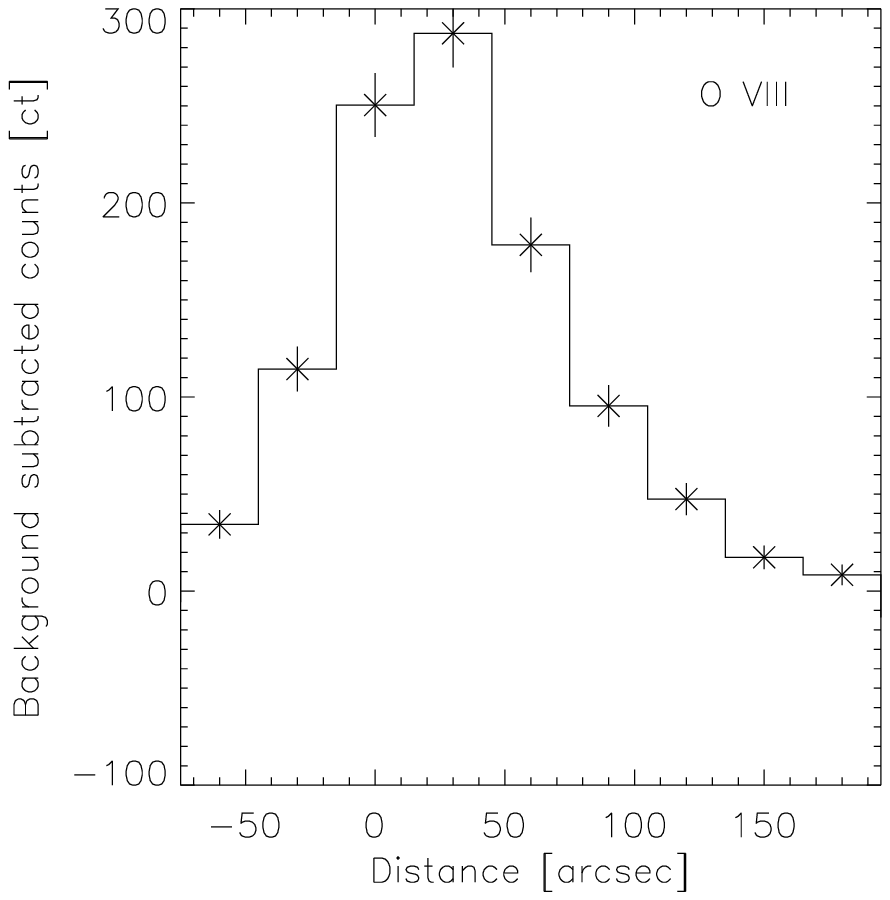}
   \includegraphics[width=5cm]{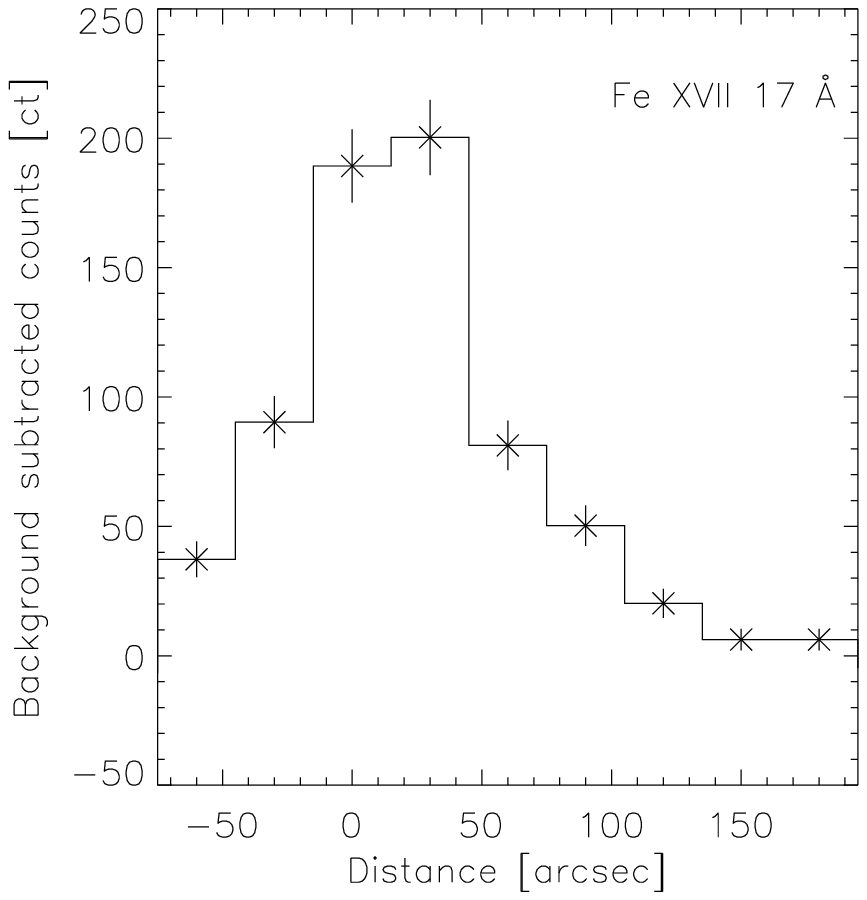}
   \caption{\ion{Ne}{X} (upper left), \ion{Fe}{XVII} at 15~\AA\ (upper right), \ion{O}{VIII} (lower left) and \ion{Fe}{XVII} at 17~\AA\ (lower right) profiles of background-subtracted counts against cross-dispersion distance for the combined \n253\ RGS1 and RGS2 data. 
   The bin at distance zero corresponds to the extraction region \lq Centre\rq.
   Negative distances are towards the north-west, and positive values correspond to areas to the south-east.
   }
   \label{fig:RGSprofilesAll}
  \end{center}
  \end{minipage}
\end{figure*}

The RGS spectra show emission in many different lines (Fig.~\ref{fig:spec}).
Especially in the region including the major axis of \n253\ (Centre) a large variety of lines from different elements can be identified.
The spectra extend from the \ion{Si}{XIV} line at the highest energy down to the \ion{C}{VI} line at the low energy end ($\lambda=6\dots$34~\AA).
All the He-like ions in this range (\ion{Si}{XIII}, \ion{Mg}{XI}, \ion{Ne}{IX}, \ion{O}{VII} and \ion{N}{VI}) and their corresponding ions in the next higher ionisation state (\ion{Si}{XIV}, \ion{Mg}{XII}, \ion{Ne}{X}, \ion{O}{VIII} and \ion{N}{VII}) can be resolved.
The iron 3$d$-2$p$ transitions around 15~\AA\ as well as the 3$s$-2$p$ lines around 17~\AA\ are detected, \ie their peak heights are more than twice the error in the wavelength bin.
Also the lines at $\sim$16.0 and $\sim$16.1~\AA\ can be clearly detected, but it is not clear whether the line at $\sim$16.0~\AA\ is from \ion{Fe}{XVIII} or from \ion{O}{VIII}.
In the range 10~\AA\ to 11.5~\AA\ there is an indication of iron lines from \ion{Fe}{XXIII} and \ion{Fe}{XXIV}.

A spectral feature that is prominent in the spectra is a broad base at the \ion{Fe}{XVII} lines at 15~\AA\ and 17~\AA\ and at the \ion{O}{VIII} and \ion{O}{VII} line positions with a width of up to $\sim$0.5~\AA. 
If this line broadening is caused by the velocity dispersion of the outflowing gas, it would imply deprojected velocities of up to $\sim$40000~km/s, corresponding to temperatures of $\sim 10^{11}$~K. 
Such high velocities are in disagreement with outflow velocities in other galaxies. 
Extreme cases show values of up to $\sim$3600~km/s \citep[e.g.\ in NGC~3079,][]{VC1994}.
A more plausible explanation for the broad base is contributions from the disc emission of \n253:
In Fig.~\ref{fig:BrightnessProfile} disc emission extends about 200\arcsec\ in dispersion direction (major axis of the galaxy) in both directions from the centre. 
This corresponds to a $\sim$0.5~\AA\ shift in wavelength and therefore can explain the observed effect.

The region north-west of the centre (NW) is strongly affected by absorption from the galactic disc that lies between the outflow and the observer.
\cite{PV2000} derive an additional absorbing column \nh\ of $1-2\times 10^{21}$~cm$^{-2}$ for this position north-west of the centre.
Therefore most of the lines are only weak or not detectable at all.
If we assume that the north-west and south-east outflow have a similar intrinsic spectrum, then absorption can fully account for the difference between regions NW and SE~1, at the same projected distance from the centre.
The strongest lines in the NW region are from \ion{Ne}{X}, \ion{Ne}{IX}, \ion{Fe}{XVII} at $\sim$15~\AA\ and $\sim$17~\AA, \ion{O}{VIII} and from the forbidden line of the \ion{O}{VII} triplet, but most of the emission originates from the disc and is smoothed out into the broad base, as mentioned above.

The regions south-east (SE~1 to SE~3) of the centre do not suffer from this absorption by the disc.
One can easily follow how the lines increase or decrease in strength as one is going away from the galactic disc south-east along the minor axis.
The lines at short wavelengths from silicon and magnesium are the first to disappear with distance from the centre.
The \ion{Ne}{X} line is seen to decrease considerably in strength south-east of the Centre region, whereas the line from the lower ionised \ion{Ne}{IX} is not affected as much and is even stronger than the \ion{Ne}{X} line in regions SE~1 and SE~2.
All the lines from iron decrease in strength except for the \ion{Fe}{XVII} lines at 17~\AA\ which grow by a factor of $\sim$1.5.
Also the \ion{O}{VIII} line increases in strength.
The \ion{O}{VII} triplet has about the same strength in region SE~1 as in the Centre region.
The lines from \ion{N}{VII} and \ion{C}{VI} increases in strength compared to the central region.

Further away from the centre, in region SE~2, the \ion{Fe}{XVII} lines at $\sim$15~\AA\ are still detectable and the \ion{O}{VIII} line is the strongest line in the spectrum.
Also the \ion{O}{VII} triplet is still strong.
The detection of all the other lines is below 2$\sigma$, even though the lines at wavelengths longer than 12~\AA\ can still be identified.

In region SE~3 only very weak lines from \ion{O}{VIII} and from the \ion{Ne}{IX} triplet remain.

Unfortunately the statistics in the spectra are not good enough to allow a quantitative spectral analysis with XSPEC.
However, several conclusions can be drawn.

Temperature estimates can be inferred from line ratios of different elements or of the same element in different ionisation states using model calculations. 
Assuming collisional ionisation equilibrium (CIE), \cite{MG1985} calculated line strengths for different elements and transitions depending on the temperatures of the plasma. 
By measuring the fluxes of two transitions in a spectrum and taking the ratio of these, the obtained value can be compared with the tables in \cite{MG1985} and the temperature of the plasma can be derived.

We used the line strength ratio between the Ly$_{\alpha}$ state of a given element and its helium-like charge state that matched the morphology of the Ly$_{\alpha}$ state in the RGS images best.
The line strengths were derived by integrating the flux of the line over the wavelength (cf.\ Table~\ref{tab:lineflux}).
We found that the derived temperature value and its error show only a weak dependence on how much of the wings of the line we include in the flux integration.
The resulting temperatures for Si, Mg, Ne and O and their variation along the outflow direction are shown in Table~\ref{tab:temperatures}.
Using only the peak height of the line however gives temperatures that are lower, except for oxygen in the regions Centre and SE~1, compared to the values shown in Table~\ref{tab:temperatures} by up to 50\%.

\begin{table}
  \begin{center}
    \caption{Temperatures of the plasma for different regions of
    the outflow of \n253\ derived from line ratios of different
    elements. For region SE~3 no temperatures could be derived as
    the lines are too weak.}
    \vspace{1em}
    \begin{footnotesize}
    \renewcommand{\arraystretch}{1.2}
    \begin{tabular}[h]{lcccc}
\hline\noalign{\smallskip}
\hline\noalign{\smallskip}
      Region & \multicolumn{4}{c}{Temperature in keV}\\
            & Si        	& Mg        	& Ne        	& O 		\\
\noalign{\smallskip}
      \hline
\noalign{\smallskip}
      NW    &           	& 0.61$\pm$0.08 & 0.51$\pm$0.08 & 0.21$\pm$0.01	\\
      Centre& 0.79$\pm$0.06 	& 0.66$\pm$0.04 & 0.43$\pm$0.02 & 0.22$\pm$0.01	\\
      SE 1  &           	& 0.46$\pm$0.04 & 0.38$\pm$0.03 & 0.21$\pm$0.01	\\
      SE 2  &           	&       	& 0.25$\pm$0.02 & 0.31$\pm$0.04	\\
\noalign{\smallskip}
      \hline \\
      \end{tabular}
      \end{footnotesize}
    \label{tab:temperatures}
  \end{center}
\end{table}

Using the line flux from an emission line (cf. Table~\ref{tab:lineflux}) and an estimate of the size of the emitting region we can derive electron densities for the nuclear region of \n253\ and the south-east outflow.
\cite{MG1985} give the line power $P^{\prime}$ normalized to the electron density for different temperatures and X-ray emission lines.
The electron density can then be derived using the formula
\begin{equation}
n_e=\sqrt{\frac{F E_{\gamma} 4 \pi d^2}{V P^{\prime}}}
\end{equation}
where $n_e$ is the electron density, $F$ the flux (in counts~s$^{-1}$~cm$^{-2}$) in an emission line with $E_{\gamma}$ from an emitting region with volume $V$ and distance $d$. The emitting region is assumed to be uniform in $n_e$.

For the central region we selected the \ion{Ne}{X} line and assumed a uniform, spherical emitting region with a radius of $\sim$160~pc.
The outflow in region SE~1 is best represented in the \ion{O}{VIII} line and we assumed a uniform, cylindrical volume with a radius of $\sim$200~pc and a height of 375~pc, the latter being confined by the extraction region.
We selected these lines because they are strong in the spectra and because we can get a good estimate for the emitting volume from the RGS images.
The resulting electron densities are $n_{\mathrm{e, nucleus}}=0.106\pm0.018$~cm$^{-3}$ and $n_{\mathrm{e, outflow}}=0.025\pm0.003$~cm$^{-3}$ for the nucleus and the outflow region, respectively.

Apart from the derivation of temperatures and electron densities, some selected emission lines can be used as diagnostic lines.

The \ion{Fe}{XVII} lines at 15~\AA\ and 17~\AA\ can be used to derive the ionising mechanism in the plasma.
In the Centre region of \n253\ as well as in the regions NW and SE~2 the line strengths indicate a predominantly collisional ionised plasma.
Region SE~1, however, shows an inverted line ratio.
Here the lines at 17~\AA\ are stronger than the lines at 15~\AA, which points at a photoionised plasma.

In general the helium-like line triplet of \ion{O}{VII} can provide the electron density, the electron temperature as well as the ionisation process \citep{PM2001}.
However, we refrained from using the \ion{O}{VII} triplet for the following reasons:
In the combined and fluxed spectra the significance is below 3~$\sigma$ for most of the spectral bins.
Also the individual lines are not clearly distinguishable from each other.
The latter is probably enhanced due to the use of the task {\tt rgsfluxer} as described above.
We also refrained from doing a simultaneous fit of the single uncombined spectra with XSPEC, as the statistics in one single spectrum are barely above $2\,\sigma$ for the strongest bin.

\subsection{RGS cross-dispersion profiles}
In the cross-dispersion profiles (Fig.~\ref{fig:RGSprofilesAll}) line emission is strongest in the Centre bin at a distance of 0\arcsec\ for \ion{Ne}{X} and \ion{Fe}{XVII} at 15~\AA. 
In the \ion{O}{VIII} and \ion{Fe}{XVII} at 17~\AA\ profile this is not the case. 
Here the SE~1 region at a distance of +30\arcsec\ is the brightest.
However the north-west half of the central bin is already affected by absorption from the disc.
When one bins the brightness profile into smaller spatial regions, the strength drops dramatically in the central bin and towards the north-west.
Towards the south-east, on the other hand, it only drops slowly.
Further to the south-east with distances larger than 45\arcsec\ the \ion{Ne}{X} and the \ion{Fe}{XVII} profiles show a large drop in brightness, whereas the \ion{O}{VIII} profile declines with a less steep gradient. 
Therefore the \ion{Ne}{X} and \ion{Fe}{XVII} emissions are more concentrated within the disc, whereas the \ion{O}{VIII} emission extends farther away from the disc.
The general behaviour that the emission from higher energy lines is more concentrated and not as extended is also seen in \m82\ \citep{RS2002}.

\begin{figure*}
   \centering
   \includegraphics[width=14cm]{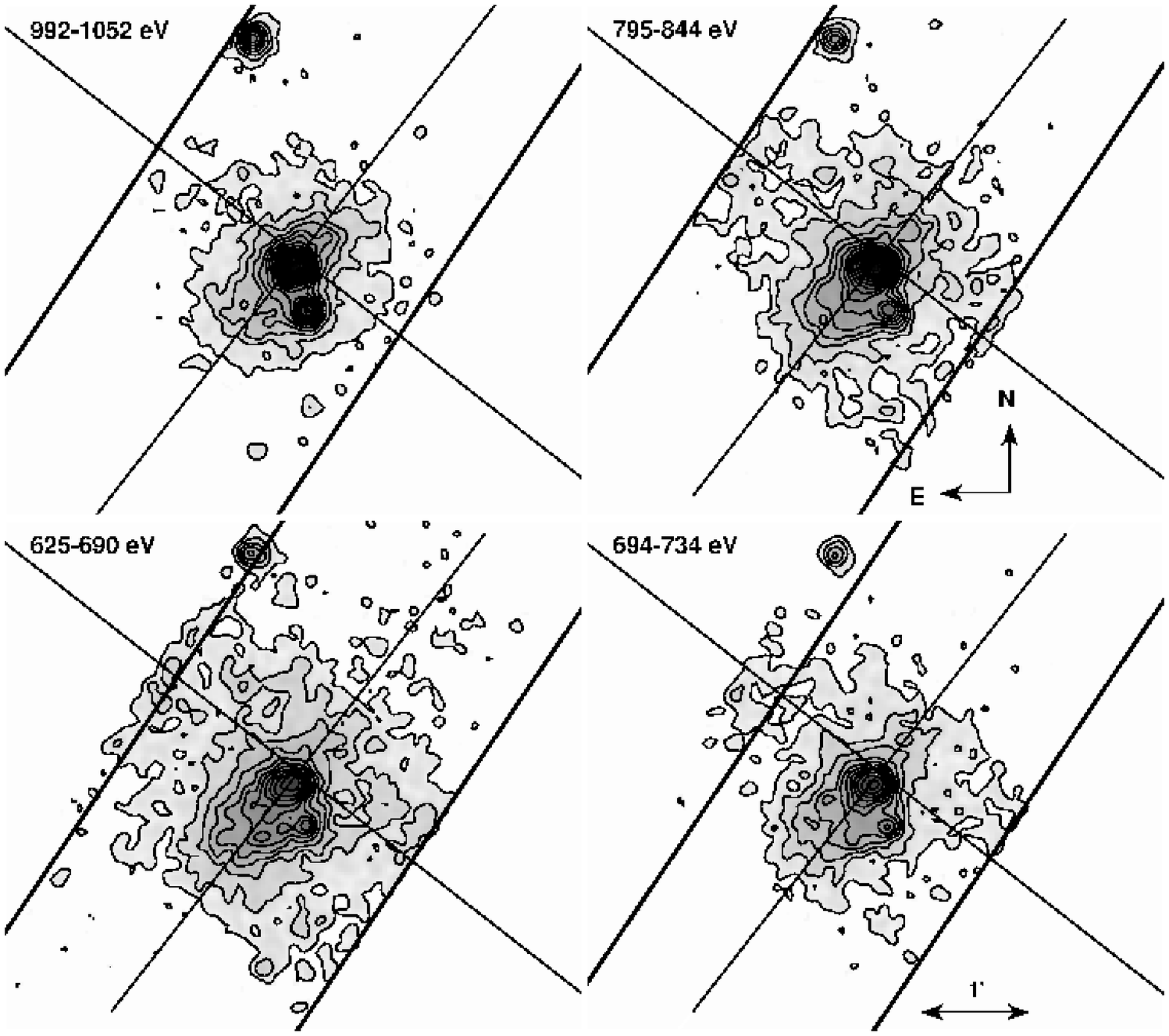}
   \caption{EPIC~PN images of \n253\ in the energy bands around the \ion{Ne}{X} (922--1052~eV), \ion{Fe}{XVII} (795--844~eV and 694--734~eV) and \ion{O}{VIII} (625--690~eV) lines.
   The images show the region of \n253\ that is covered by the RGS images (Fig.~\ref{fig:RGSNeFeO}).
   The thin black lines mark the major (north-east to south-west) and minor axis (north-west to south-east) of \n253.
   The area within the thick black lines gives the extraction region in the corresponding RGS line image.
   Black contours indicate 2$\sigma$, 3$\sigma$, 4$\sigma$,~\dots above the background.
   The images were smoothed using a gaussian with a FWHM of 6\arcsec.
   }
   \label{fig:EPICNeFeO}
\end{figure*}

\subsection{RGS images}
The RGS images allow us to derive additional information about the spatial composition of the outflow.
But before we describe the images in more detail we need to point out that these images are dominated by low statistics.
We therefore restrict our results and conclusions on the regions with good statistics.

The strong \ion{O}{VIII} emission clearly traces the outflow.
It reaches out to $\sim$750~pc projected distance along the south-east minor axis and has an extent perpendicular to that of $\sim$400~pc.
Because of the high absorption, emission from the farther side of the disc is not detected.
With the RGS spatial resolution, the outflow in \ion{O}{VIII} does not show limb brightening.
The emission is strongest close to its central axis and its intensity decreases towards the border.
This suggests that the emitting ions are not concentrated on the borders of the outflow where the outflow gas interacts with the surrounding material, but that this emission comes directly from the outflowing gas.
The extent to the south-east is less than the one in the EPIC images where the bright outflow emission can be followed to $\sim$1.3~kpc (Fig.~\ref{fig:EPICNeFeO}), due both to the sensitivity and to the smaller energy band ($\Delta \lambda=$1.87~\AA\ vs. 0.37~\AA) that was used to extract the images for the RGS.
The emission from \ion{Ne}{X} is clearly concentrated in the disc and nucleus and it gives no contribution to the outflow.
The images in the iron lines both show the strongest emission south-east of the nucleus and the emission is slightly extended along the outflow direction.
However they do not trace the outflow morphology as seen in the \ion{O}{VIII} line.
This is not surprising, because the excitation cross sections for the iron and \ion{O}{VIII} lines 
have a different temperature dependence. 
Due to the work that is performed when the outflow expands against the pressure of the ambient medium,
as well as to the divergence of the flow in the cone perpendicular to the disc, the temperature, or more precisely the kinetic energy of the electrons, should decrease \citep[e.g.][]{BS1999} with height $z$ above the disc, explaining the relative increase in emission of \ion{O}{VIII} to iron.

\subsection{EPIC-PN images}
We find that the EPIC~PN narrowband images (Fig.~\ref{fig:EPICNeFeO}) are all affected by the redistribution effect of the detector, \ie the image is contaminated by events with higher energies from outside the energy filter boundaries.
Especially very bright sources like the central source of the galaxy contribute strongly to this effect and all images will therefore show these sources.
The bright nuclear source is clearly visible in all of the EPIC~PN images.
Also the bright source $\sim$0.5\arcmin\ south-west of the nucleus is clearly visible in all EPIC~PN images.
For the RGS images the latter is only true for the \ion{Fe}{XVII} at 15~\AA\ image.
This clearly shows the advantage, namely the far better energy resolution, of the RGS images compared to the EPIC~PN narrowband images.

With respect to the limb brightening of the outflow emission, the image in the energy range 694--734~eV (including the \ion{Fe}{XVII} at 17~\AA\ line) indicates a morphology that could result from a limb brightened outflow at a distance of $\sim$0.7' away from the galactic centre.
Also in the energy range 625--690~eV (including the \ion{O}{VIII} line) there is an indication of this morphology at a distance of about 1.1'.
The corresponding RGS image unfortunately has too low statistics to confirm this.
The other images give no indication of limb brightening.

\subsection{EPIC-PN brightness profiles}
\begin{figure*}
   \centering
   \includegraphics[width=\textwidth]{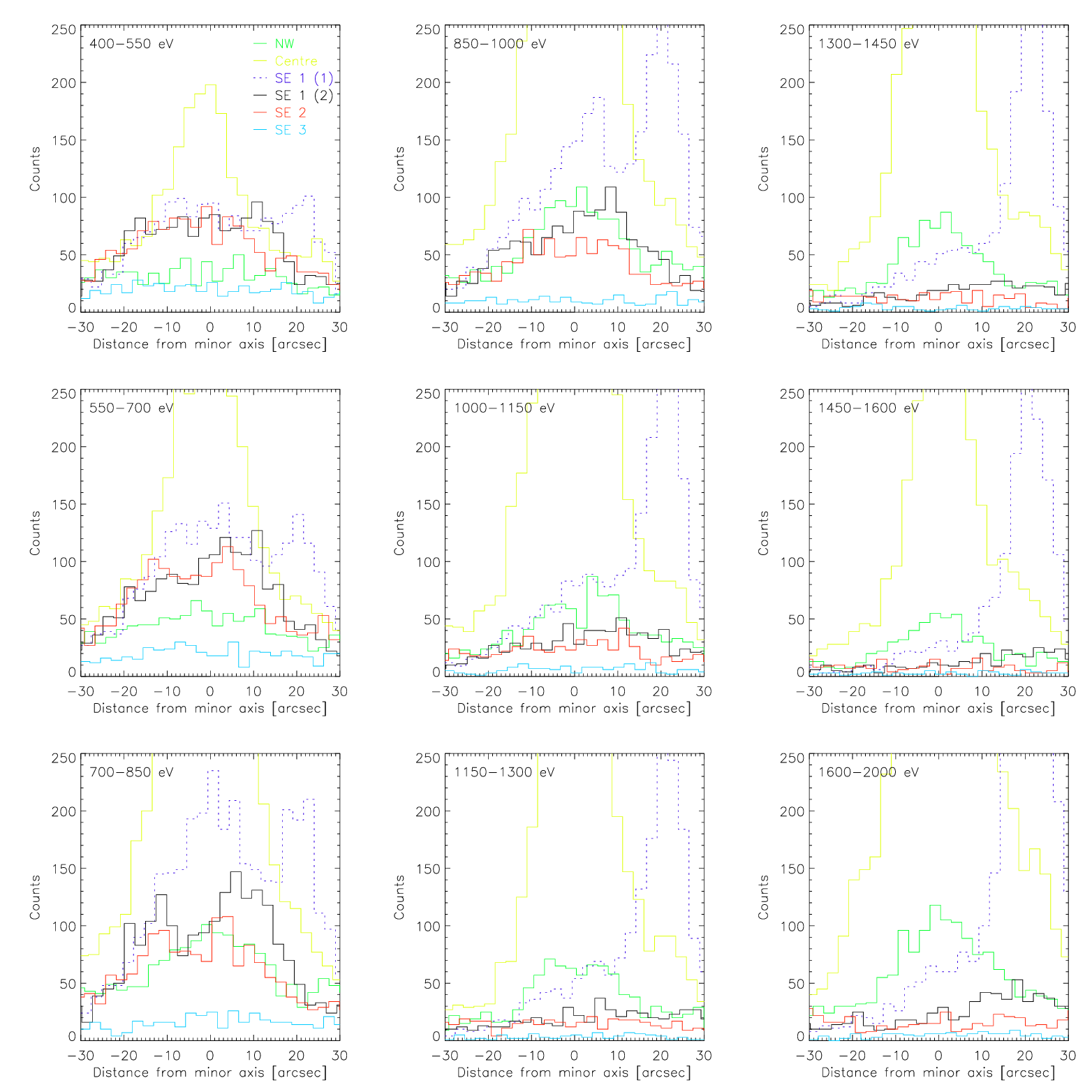}
   \caption{EPIC~PN brightness profiles of the outflow region for the RGS extraction regions sorted by energy band.
   Note that region SE~1 was split into the two regions 'SE~1 (1)' and 'SE~1 (2)' with a width of 15\arcsec\ in cross-dispersion direction of the RGS.
   The abscissa gives the distance from the minor axis of the galaxy in arcsec.
   Positive values are to the south-west, negative values to the north-east.
   The presented histograms consist of raw detector counts, \ie the background and the detector characteristics, like quantum efficiency or effective area, were not taken into account.
   }
   \label{fig:PN_profiles}
\end{figure*}

In the EPIC-PN brightness profiles (Fig.~\ref{fig:PN_profiles}) the peak in region SE~1~(1) that is caused by the outflow emission coincides always with the centre.
A double peak with a central depression is visible in SE~1~(2) and SE~2 at 700--1000~eV (covering also the iron lines).
This may point at a limb brightened outflow in this region.
An indication of this structure in region SE~2 at 550--850~eV could be a redistribution effect of the detector (see above).
At other energies no indication of limb brightening can be seen. 
The profiles show either a flat plateau or a hump peaking around the centre of the outflow.
This confirms the findings of the RGS images.
Starting from 700~eV region NW comes out and it is stronger than the two SE~1 regions together in energies above 1150~eV.
The strong peak in the region SE~1~(1) (blue, dotted line) at $\sim$+20\arcsec\ distance is the bright point source south-west of the nucleus and not part of the outflow region.

To rule out that our findings are affected by \xmm's spatial resolution we compared brightness profiles, extracted with the same regions and energy bands from \chandra\ observation 3931, to the EPIC~PN profiles.
There the double peak is clearly detected in region SE~1~(2) between 700 and 1000~eV, possibly also up to 1150~eV.
It is not visible in regions SE~1~(1) or SE~2.
The profiles in region NW show the same behaviour as in the EPIC-PN data.
So taking the differences in the instruments into account \xmm\ and \chandra\ give a consistent picture.


\section{Discussion}
\subsection{Line ratios and temperatures}
The extracted RGS spectra of the outflow along the minor axis of the galaxy show emission lines from many ions in different ionisation states: the Ly$_{\alpha}$ lines from Si, Mg, Ne, O, N and also their helium-like charge states.
Additionally we see emission lines from \ion{Fe}{XVII} and \ion{Fe}{XVIII}.

With increasing distance from the nuclear region, the relative flux in the \ion{O}{VIII} line intensifies compared to the flux in \ion{Fe}{XVII}, \ie the line flux at longer wavelengths increases.
The same effect can also be seen in the line flux ratio of O~VII to O~VIII.
For the oxygen line ratio this implies that the temperature decreases; the gas is cooling as it flows away from the nucleus.
This change in temperature would also affect the excitation of \ion{Fe}{XVII}. 
Since both elements, oxygen and iron, are from the same kind of sources, i.e.\ type~II~SNe, it seems unlikely that the distribution is different, unless there is a very different clumpiness in O and Fe initially.
Therefore a change in the oxygen to iron line flux is more likely due to the change in temperature than due to a different radial abundance profile.

The line strengths of the \ion{Fe}{XVII} lines at 15~\AA\ compared to the one at 17~\AA\ indicates a predominantly collisional ionised plasma.
Region SE~1, however, shows an inverted line ratio.
There, the flux in the \ion{Fe}{XVII} lines at 17~\AA\ is enhanced compared to the lines at 15~\AA.
This points at a photoionised plasma, but no strong photoionising sources were detected in the vicinity.
There are two additional alternatives to create this line ratio:

\noindent A) the plasma is highly underionised compared to the ionising electrons.
The time since the heating of the plasma was too short to reach an equilibrium state.
In these so called underionised plasmas inner-shell ionisation is highly operational \citep{K2006} and leads to an enhanced 17.10~\AA\ flux from \ion{Fe}{XVII} \citep{DB2002}. 
Examples that show underionised plasmas are supernova remnants like N132D \citep{BR2001} and Dem L71 \citep{HB2003}.

\noindent B) the plasma is overionised. 
It can be produced in fast adiabatic cooling of hot ($T\sim 10^8$~K) and almost completely ionised gas expanding out of a superbubble \citep{BS1999}.
The ionisation cross sections as well as the recombination cross sections are different for all ions, so when a shock propagates through the outflow region, the ionisation of different elements and different ionisation stages occurs at different time scales for each.
As the expansion of the wind occurs on much shorter time scales than radiative recombination, the highly ionised atomic states remain frozen-in.
These states then recombine at a later time further away from the nucleus.
This is called delayed recombination.
In both regions, Centre and SE~1, we see emission from \ion{Fe}{XVIII}, which is produced in the hot starburst region of the galaxy.
As the wind breaks out of the nuclear superbubble and expands into the lower density medium along the minor axis of the galaxy it cools mostly adiabatically.
However the \ion{Fe}{XVIII} in the wind recombines on a larger time scale than the wind fluid, so we have \ion{Fe}{XVIII} in abundance although the electron temperature would preferentially admit \ion{Fe}{XVII}.
When the \ion{Fe}{XVIII} ions then recombine to \ion{Fe}{XVII}, the states of \ion{Fe}{XVII} which decay via the 17~\AA\ lines are preferably populated compared to the ones which decay via the 15~\AA\ lines.
This enhances the flux at 17~\AA\ and could lead to the misleading line ratio in this region.

Temperatures derived from line ratios of individual elements are in the range of $0.21\pm0.01$ to $0.79\pm0.06$~keV.
This range is larger than the one found by \cite{SH2000} using \chandra\ ($0.46^{+0.11}_{-0.10}-0.66^{+0.10}_{-0.08}$~keV).
However \cite{SH2000} used only a single temperature MEKAL hot plasma model for each extraction region.
Given the complexity of the X-ray producing mechanisms this approach is definitely too simple.
Our derived temperature range is closer to that given by \cite{PR2001} from \xmm\ EPIC spectra, who used three MEKAL models (0.15, 0.53 and 0.94~keV).
The approach in this paper to derive temperatures tries to account for the high complexity of the outflow, without the claim to be complete, by using actually one MEKAL model for each element, and by taking advantage of the superior energy resolution of the RGS. 
This approach gives already almost the whole measured energy range just for the Centre region, whereas \cite{SH2000} obtain this range of temperatures by using several different extraction regions along the outflow.
Looking at the temperatures of just one extraction region we see that different elements give different temperatures.
This could be due to sampling of different regions along the line of sight, but also due to delayed recombination.
The latter is a true non-equilibrium ionisation (NEI) situation and the result would be measuring different temperatures for different ions.
All of the above temperatures were derived by the use of models that assume CIE for the emitting gas.
This is not necessarily the case for an outflow.
To avoid this assumption we would have to fit the spectra to NEI models \citep[e.g.\ from][]{BS1999}.
These models entail a higher level of complexity as they depend crucially on a detailed 
hydrodynamical model of the outflow. Therefore they are sensitive to the time evolution 
of the wind, and as such unique. In other words, NEI models would enable us to fully exploit 
the information buried in the X-ray spectra, since one would calculate a series of models 
with different boundary conditions, and then derive suitably binned synthetic spectra for 
fitting the observational results. Thus a satisfactory fit would determine within certain limits 
the physical parameters of the outflow.

The Fe lines could in principle be used as well to derive temperatures.
However the emission from \ion{Fe}{XVII} at $\sim$15$\,$\AA\ (2p-3d) and at $\sim$17$\,$\AA\ (2p-3s) are affected by delayed recombination of \ion{Fe}{XVIII}, which contradicts the CIE assumption.
Therefore the results could be misleading and are not shown in Table~\ref{tab:temperatures}. 
However from the fact that the \ion{Fe}{XVII} at 15~\AA\ lines are present and strong in the spectra, the temperature has to be above 0.22~keV in the regions where the Fe~L shell originates from, otherwise \ion{Fe}{XVII} line formation is inefficient \citep{RM1985}.
Furthermore calculations show that above 0.60~keV \ion{Fe}{XVIII} is strong while \ion{Fe}{XVII} is weak. 
This is not the case in any of the regions where we obtained spectra from, therefore we expect that kT=$0.2\dots 0.6$~keV

The RGS images give another indication of the cause of the large temperature spread. 
The different distribution of the events for each emission line strongly points to the fact that plasmas in different regions of space are responsible for the line emission. 
It is not surprising that these plasmas then do not have the same temperature and that the temperatures for each element are different.
The time scale for attaining pressure and hence temperature equilibrium in the 
starburst region is of the order of the sound crossing time.
If we compare this to the average time interval between supernova explosions, 
we can assess the smoothness of the temperature distribution. 
A rate of $0.05 \, {\rm yr}^{-1}$ has been quoted for \n253\ \citep{CP1992}. 
A supernova blast wave expanding within a hot tenuous gas has a low Mach number,
so that the sound crossing time, $\tau_{sc}$, in the starburst region is roughly the time scale for 
the shock wave to reheat the hot gas. Adopting a value of $20$\arcsec\ or $d=250$~pc 
and a temperature of $0.6$~keV or $6.6 \times 10^6$~K for the central region yields a value of 
$\tau_{sc} \sim d/c_s \simeq 8.2 \times 10^5 \, {\rm yr}$, which is much larger than the 
interval of $20$ years between successive explosions, even if we decrease the supernova rate
by a factor of $10 - 100$ for the starburst region. Therefore temperature inhomogeneities
in the starburst region and hence variations in the ionisation stages are to be expected.
As the flow moves away from the disc sources, the temperature structure should however 
become smoother with time when the flow time becomes larger than the sound crossing time $\tau_{sc}$.

\cite{SH2000} argue that the spectral variation in \chandra\ spectra \lq along and between the northern and southern outflow cones is due to variations in the absorption column and not due to significant temperature variations along the outflow\rq. 
A higher absorption column is certainly the case for the NW region, resulting in too high temperature values.
Although from the analysis of the \xmm\ RGS spectra we cannot draw conclusions regarding the absorbing column, we find that there is a significant temperature variation along the outflow.
The temperatures for Mg and Ne are decreasing with distance to the south-east from the centre.
The temperature of oxygen seems to be constant, except for the value in region SE~2.

\subsection{The morphology of the outflow}
The only RGS image that clearly shows the outflow geometry is the one in the \ion{O}{VIII} line.
The shape of the south-east outflow matches roughly the truncated cone with an opening angle of $\sim$26\deg\ that is seen in \chandra\ and H$_{\alpha}$ data \citep{SH2000}, though the angle in the \ion{O}{VIII} image appears to be slightly smaller.
Also this emission is not limb brightened.
The image suggests that the outflow cone is filled with clumpy distributed \ion{O}{VIII}.
If this is correct, then we see the emission from the hot wind fluid itself for the first time.
The clumpiness may be caused by mass-loading, \ie the turbulent process of mixing in ambient ISM and infalling material.
This detection of the wind fluid however is in contrast to the statement from \cite{SS2000} that the wind is too thin to emit efficiently enough to be detected.

In \xmm\ EPIC observations by \cite{PR2001} and in \chandra\ observations by \cite{SH2000} the outflow shows a limb brightened morphology.
In our analysis we find limb brightening in the \ion{Fe}{XVII} at 17~\AA\ RGS images.
The comparison between the different, though with lower spectral resolution, PN images and brightness profiles also shows an energy dependence of limb brightening and a clear detection of the emission from the outflowing gas at the energies below 700~eV.

\cite{SS2000} found from simulations that low opening angles for outflows are only possible when a thick galactic disc is present.
This produces opening angles from a few degrees when the wind starts to blow out of its superbubble
and then increases up to values of $\sim 60\deg$, depending on boundary conditions.
Thin-disc models typically show opening angles of $\sim 90\deg$.
However these simulations assumed an ISM distribution in rotating hydrostatic equilibrium with the gravitational field.
Simulations by \cite{AB2004,AB2005} show that the ISM in the disc has a highly complex structure.
This affects the break-out dynamics of a superbubble via density and pressure gradients and also via "holes" from previous outflows that have not closed yet.
Conclusions about the thickness of the disc and its collimating effect are hence difficult.

\subsection{Outlook}
In a forthcoming paper we will analyse the \xmm\ EPIC data of the outflow and the halo emission with the NEI model from \cite{BS1999}, which will then also consider the ionisation history of the outflowing gas from the nucleus into the halo.
This will give further information on the composition, temperature and dynamical structure of the outflow.


\begin{acknowledgements}
The \xmm\ project is supported by the Bundesministerium f\"ur Wirtschaft und Technologie/Deutsches Zentrum
f\"ur Luft- und Raumfahrt (BMWI/DLR, FKZ 50 OX 0001), the Max-Planck Society and the Heidenhain-Stiftung.
This research has made use of the NASA/IPAC Extragalactic Database (NED) which is operated by the Jet Propulsion Laboratory, California Institute of Technology, under contract with the National Aeronautics and Space Administration.
MB acknowledges support from the BMWI/DLR, FKZ 50 OR 0405.
GT acknowledges the financial contribution  from ASI under contract ASI-INAF I/023/05/0.
Finally we thank the referee for helpful comments that allowed us to improve the paper.
\end{acknowledgements}

\bibliographystyle{aa}
\bibliography{papers}

\begin{thebibliography}{28}
\expandafter\ifx\csname natexlab\endcsname\relax\def\natexlab#1{#1}\fi

\bibitem[{{Behar} {et~al.}(2001){Behar}, {Rasmussen}, {Griffiths}, {Dennerl},
  {Audard}, {Aschenbach}, \& {Brinkman}}]{BR2001}
{Behar}, E., {Rasmussen}, A.~P., {Griffiths}, R.~G., {et~al.} 2001, \aap, 365,
  L242

\bibitem[{{Breitschwerdt} \& {Schmutzler}(1999)}]{BS1999}
{Breitschwerdt}, D. \& {Schmutzler}, T. 1999, \aap, 347, 650

\bibitem[{{Colina} \& {Perez-Olea}(1992)}]{CP1992}
{Colina}, L. \& {Perez-Olea}, D. 1992, \mnras, 259, 709

\bibitem[{{de Avillez} \& {Breitschwerdt}(2004)}]{AB2004}
{de Avillez}, M.~A. \& {Breitschwerdt}, D. 2004, \aap, 425, 899

\bibitem[{{de Avillez} \& {Breitschwerdt}(2005)}]{AB2005}
{de Avillez}, M.~A. \& {Breitschwerdt}, D. 2005, \aap, 436, 585

\bibitem[{{den Herder} {et~al.}(2001){den Herder}, {Brinkman}, {Kahn},
  {Branduardi-Raymont}, {Thomsen}, {Aarts}, {Audard}, {Bixler}, {den Boggende},
  {Cottam}, {Decker}, {Dubbeldam}, {Erd}, {Goulooze}, {G{\"u}del}, {Guttridge},
  {Hailey}, {Janabi}, {Kaastra}, {de Korte}, {van Leeuwen}, {Mauche},
  {McCalden}, {Mewe}, {Naber}, {Paerels}, {Peterson}, {Rasmussen}, {Rees},
  {Sakelliou}, {Sako}, {Spodek}, {Stern}, {Tamura}, {Tandy}, {de Vries},
  {Welch}, \& {Zehnder}}]{HB2001}
{den Herder}, J.~W., {Brinkman}, A.~C., {Kahn}, S.~M., {et~al.} 2001, \aap,
  365, L7

\bibitem[{{Doron} \& {Behar}(2002)}]{DB2002}
{Doron}, R. \& {Behar}, E. 2002, \apj, 574, 518

\bibitem[{{Fabbiano} \& {Trinchieri}(1984)}]{FT1984}
{Fabbiano}, G. \& {Trinchieri}, G. 1984, \apj, 286, 491

\bibitem[{{Forbes} {et~al.}(2000){Forbes}, {Polehampton}, {Stevens}, {Brodie},
  \& {Ward}}]{FP2000}
{Forbes}, D.~A., {Polehampton}, E., {Stevens}, I.~R., {Brodie}, J.~P., \&
  {Ward}, M.~J. 2000, \mnras, 312, 689

\bibitem[{{Jansen} {et~al.}(2001){Jansen}, {Lumb}, {Altieri}, {Clavel}, {Ehle},
  {Erd}, {Gabriel}, {Guainazzi}, {Gondoin}, {Much}, {Munoz}, {Santos},
  {Schartel}, {Texier}, \& {Vacanti}}]{JL2001}
{Jansen}, F., {Lumb}, D., {Altieri}, B., {et~al.} 2001, \aap, 365, L1

\bibitem[{{Koribalski} {et~al.}(2004){Koribalski}, {Staveley-Smith}, {Kilborn},
  {Ryder}, {Kraan-Korteweg}, {Ryan-Weber}, {Ekers}, {Jerjen}, {Henning},
  {Putman}, {Zwaan}, {de Blok}, {Calabretta}, {Disney}, {Minchin}, {Bhathal},
  {Boyce}, {Drinkwater}, {Freeman}, {Gibson}, {Green}, {Haynes}, {Juraszek},
  {Kesteven}, {Knezek}, {Mader}, {Marquarding}, {Meyer}, {Mould}, {Oosterloo},
  {O'Brien}, {Price}, {Sadler}, {Schr{\"o}der}, {Stewart}, {Stootman}, {Waugh},
  {Warren}, {Webster}, \& {Wright}}]{KS2004}
{Koribalski}, B.~S., {Staveley-Smith}, L., {Kilborn}, V.~A., {et~al.} 2004,
  \aj, 128, 16

\bibitem[{{Kosenko}(2006)}]{K2006}
{Kosenko}, D.~I. 2006, \mnras, 369, 1407

\bibitem[{{Mewe} {et~al.}(1985){Mewe}, {Gronenschild}, \& {van den
  Oord}}]{MG1985}
{Mewe}, R., {Gronenschild}, E.~H.~B.~M., \& {van den Oord}, G.~H.~J. 1985,
  \aaps, 62, 197

\bibitem[{{Phillips} {et~al.}(1999){Phillips}, {Mewe}, {Harra-Murnion},
  {Kaastra}, {Beiersdorfer}, {Brown}, \& {Liedahl}}]{PM1999}
{Phillips}, K.~J.~H., {Mewe}, R., {Harra-Murnion}, L.~K., {et~al.} 1999, \aaps,
  138, 381

\bibitem[{{Pietsch} {et~al.}(2001){Pietsch}, {Roberts}, {Sako}, {Freyberg},
  {Read}, {Borozdin}, {Branduardi-Raymont}, {Cappi}, {Ehle}, {Ferrando},
  {Kahn}, {Ponman}, {Ptak}, {Shirey}, \& {Ward}}]{PR2001}
{Pietsch}, W., {Roberts}, T.~P., {Sako}, M., {et~al.} 2001, \aap, 365, L174

\bibitem[{{Pietsch} {et~al.}(2000){Pietsch}, {Vogler}, {Klein}, \&
  {Zinnecker}}]{PV2000}
{Pietsch}, W., {Vogler}, A., {Klein}, U., \& {Zinnecker}, H. 2000, \aap, 360,
  24

\bibitem[{{Porquet} {et~al.}(2001){Porquet}, {Mewe}, {Dubau}, {Raassen}, \&
  {Kaastra}}]{PM2001}
{Porquet}, D., {Mewe}, R., {Dubau}, J., {Raassen}, A.~J.~J., \& {Kaastra},
  J.~S. 2001, \aap, 376, 1113

\bibitem[{{Puche} {et~al.}(1991){Puche}, {Carignan}, \& {van Gorkom}}]{PC1991}
{Puche}, D., {Carignan}, C., \& {van Gorkom}, J.~H. 1991, \aj, 101, 456

\bibitem[{{Read} \& {Stevens}(2002)}]{RS2002}
{Read}, A.~M. \& {Stevens}, I.~R. 2002, \mnras, 335, L36

\bibitem[{{Rugge} \& {McKenzie}(1985)}]{RM1985}
{Rugge}, H.~R. \& {McKenzie}, D.~L. 1985, \apj, 297, 338

\bibitem[{{Stevens} {et~al.}(2003){Stevens}, {Read}, \&
  {Bravo-Guerrero}}]{SR2003}
{Stevens}, I.~R., {Read}, A.~M., \& {Bravo-Guerrero}, J. 2003, \mnras, 343, L47

\bibitem[{{Strickland} {et~al.}(2000){Strickland}, {Heckman}, {Weaver}, \&
  {Dahlem}}]{SH2000}
{Strickland}, D.~K., {Heckman}, T.~M., {Weaver}, K.~A., \& {Dahlem}, M. 2000,
  \aj, 120, 2965

\bibitem[{{Strickland} \& {Stevens}(2000)}]{SS2000}
{Strickland}, D.~K. \& {Stevens}, I.~R. 2000, \mnras, 314, 511

\bibitem[{{Str{\"u}der} {et~al.}(2001){Str{\"u}der}, {Briel}, {Dennerl},
  {Hartmann}, {Kendziorra}, {Meidinger}, {Pfeffermann}, {Reppin}, {Aschenbach},
  {Bornemann}, {Br{\"a}uninger}, {Burkert}, {Elender}, {Freyberg}, {Haberl},
  {Hartner}, {Heuschmann}, {Hippmann}, {Kastelic}, {Kemmer}, {Kettenring},
  {Kink}, {Krause}, {M{\"u}ller}, {Oppitz}, {Pietsch}, {Popp}, {Predehl},
  {Read}, {Stephan}, {St{\"o}tter}, {Tr{\"u}mper}, {Holl}, {Kemmer}, {Soltau},
  {St{\"o}tter}, {Weber}, {Weichert}, {von Zanthier}, {Carathanassis}, {Lutz},
  {Richter}, {Solc}, {B{\"o}ttcher}, {Kuster}, {Staubert}, {Abbey}, {Holland},
  {Turner}, {Balasini}, {Bignami}, {La Palombara}, {Villa}, {Buttler},
  {Gianini}, {Lain{\'e}}, {Lumb}, \& {Dhez}}]{SB2001}
{Str{\"u}der}, L., {Briel}, U., {Dennerl}, K., {et~al.} 2001, \aap, 365, L18

\bibitem[{{Tanaka} {et~al.}(2005){Tanaka}, {Sugiho}, {Kubota}, {Makishima}, \&
  {Takahashi}}]{TS2005}
{Tanaka}, T., {Sugiho}, M., {Kubota}, A., {Makishima}, K., \& {Takahashi}, T.
  2005, \pasj, 57, 507

\bibitem[{{Turner} {et~al.}(2001){Turner}, {Abbey}, {Arnaud}, {Balasini},
  {Barbera}, {Belsole}, {Bennie}, {Bernard}, {Bignami}, {Boer}, {Briel},
  {Butler}, {Cara}, {Chabaud}, {Cole}, {Collura}, {Conte}, {Cros}, {Denby},
  {Dhez}, {Di Coco}, {Dowson}, {Ferrando}, {Ghizzardi}, {Gianotti}, {Goodall},
  {Gretton}, {Griffiths}, {Hainaut}, {Hochedez}, {Holland}, {Jourdain},
  {Kendziorra}, {Lagostina}, {Laine}, {La Palombara}, {Lortholary}, {Lumb},
  {Marty}, {Molendi}, {Pigot}, {Poindron}, {Pounds}, {Reeves}, {Reppin},
  {Rothenflug}, {Salvetat}, {Sauvageot}, {Schmitt}, {Sembay}, {Short},
  {Spragg}, {Stephen}, {Str{\"u}der}, {Tiengo}, {Trifoglio}, {Tr{\"u}mper},
  {Vercellone}, {Vigroux}, {Villa}, {Ward}, {Whitehead}, \& {Zonca}}]{TA2001}
{Turner}, M.~J.~L., {Abbey}, A., {Arnaud}, M., {et~al.} 2001, \aap, 365, L27

\bibitem[{{van der Heyden} {et~al.}(2003){van der Heyden}, {Bleeker},
  {Kaastra}, \& {Vink}}]{HB2003}
{van der Heyden}, K.~J., {Bleeker}, J.~A.~M., {Kaastra}, J.~S., \& {Vink}, J.
  2003, \aap, 406, 141

\bibitem[{{Veilleux} {et~al.}(1994){Veilleux}, {Cecil}, {Bland-Hawthorn},
  {Tully}, {Filippenko}, \& {Sargent}}]{VC1994}
{Veilleux}, S., {Cecil}, G., {Bland-Hawthorn}, J., {et~al.} 1994, \apj, 433, 48

\end{thebibliography}

\end{document}